\documentclass[prd,a4paper,superscriptaddress,nofootinbib,10pt]{revtex4} 
\usepackage{tikz}
\usetikzlibrary{decorations.markings}
\usetikzlibrary{snakes}
\usepackage{graphicx}
\usepackage{amssymb}
\usepackage{amsmath}
\usepackage{lscape}
\usepackage{mathrsfs}
\usepackage{textcomp}
\usepackage{epsfig}
\usepackage{slashed}
\usepackage{xcolor}
\usepackage{bbold}
\renewcommand{\d}{\mathrm{d}}
\newcommand{\nn}{\nonumber}

\begin{document}

\begin{flushleft}
\texttt{ZTF-EP-21-07}

\texttt{RBI-ThPhys-2021-41}
\end{flushleft}

\title{Propagation of spinors on a noncommutative spacetime:\\ 
equivalence of the formal and the effective approach}


\author{Marija Dimitrijevi\'{c} \'{C}iri\'{c}}
\email{dmarija@ipb.ac.rs}
\affiliation{ Faculty of Physics, University of Belgrade,  Studentski trg 12, 11000 Beograd, Serbia}

\author{Nikola Konjik}
\email{konjik@ipb.ac.rs}
\affiliation{ Faculty of Physics, University of Belgrade,  Studentski trg 12, 11000 Beograd, Serbia}

\author{Andjelo Samsarov}
\email{asamsarov@irb.hr}
\affiliation{Rudjer Bo\v{s}kovi\'c Institute, Bijeni\v cka  c.54, HR-10002 Zagreb, Croatia}

\date{\today}

\begin{abstract}

Some noncommutative (NC) theories posses a certain type of dualities that are implicitly built within their structure.
In this paper  we establish still another example of this kind, and we do this perturbatively in the first order of the Seiberg-Witten expansion. More precisely, we show that  red a particular model of  noncommutative $U(1)_\star $ gauge  field coupled to a  NC scalar field and to a classical geometry of the Reissner–Nordstr\" om (RN) type is to a first order in deformation  completely equivalent at the level of equations of motion to the commutative $ U(1)$
gauge theory coupled to a  commutative scalar field and to a classical geometry background, different from the starting RN background. The new (effective) metric is
 obtained from the RN metric  by switching on   an additional nonvanishing $r-\phi$ component. Using this  first order  duality between two theories and physical systems they describe, we formulate an effective approach to studying a dynamics of spin $\frac{1}{2}$
fields on the curved background of RN type with an abiding noncommutative structure.
As opposed to that, we also  investigate  in a more formal way a  dynamics of spin $\frac{1}{2}$ fields,  and we do this perturbatively, within a first order in deformation parameter,  by studying a  semiclassical  theory which describes the NC $U(1)_\star $ gauge field coupled to NC spin $\frac{1}{2}$ field and also coupled to gravity, which is however treated classically. 
 Upon utilising the Seiberg-Witten (SW) map in order to write the NC spinor  and NC gauge fields in terms of their corresponding commutative degrees of freedom, we find that the equation of motion for the fermion field  obtained within the formal approach exactly coincides with the equation of motion obtained within the effective approach that utilises   first order   noncommutative duality. Therefore, linearized equations of motion for a spinor field in SW expansion turn out to be the same as equations of motion in a perturbed metric.
We then use these results to analyze the problem of stability of  solutions of the equations of motion and the associated issue of
 superradiance,  as related to fermions in RN spacetime with an all-pervasive noncommutative structure.
\end{abstract}

\maketitle

\section{Introduction}

Many distinct approaches to a unification of quantum mechanics  with gravity in the ultraviolet sector  point toward the existence of absolute minimal length scale, whose very presence puts a lower bound to the minimal possible resolution of space as its intrinsic property.
As a consequence, one of the  cornerstones of  quantum mechanics, the Heisenberg uncertainty relations, start to call for a revision, resulting in a generalized uncertainty
principle \cite{kempf-mangano},\cite{gup-witten}, which is also suggested by perturbative string theory \cite{amati-veneziano},\cite{gross-mende}, quantum gravity \cite{garay},\cite{carlo}    and black hole physics \cite{maggiore}. Moreover, in loop quantum gravity  a process of quantization gives rise to the area and volume operators which have discrete spectra, whose lowest possible eigenvalues are being proportional to the square and cube of the Planck length, respectively \cite{rovelli},\cite{thiemann}. This, together with the generalized uncertainty principle, implies the existence  of  a minimum uncertainty in position \cite{maggiore},\cite{amati-veneziano1},\cite{hossenfelder},\cite{gup-nicolini}.

One of the more known patterns to  implement a minimal length scale in quantum mechanics, quantum field theory (QFT) and gravity is provided by  the frame of noncommutative (NC) geometry,
which is characterized by  the fact that the spacetime coordinates get raised to the level of the operators and thus generally fail to commute \cite{ahluwalia,dopl-fred,SW,nekrasov,szabo}.
The idea of noncommutativity of spacetime was first clearly articulated by Snyder \cite{snyder} and expounded further in terms of geometric notions by Alain Connes \cite{connes1},\cite{connes2}. Correspondingly, a possibility to observe the consequences of spacetime noncommutativity and the existence of a minimal length scale led 
to the intensive study of noncommutative versions of quantum mechanics, QFT and gravity   with the aim of revising the standard theories as diverse as the gauge theory, particle physics  and the Quantum Hall effect (QHE), thermodynamics and the  black-hole physics and cosmology, so that they may keep the track with and accommodate the eventual novel features  into their framework \cite{Chaichian:1998kp,Hayakawa:1999zf,sheikh2,arefeva,Micu:2000xj,jssw,Chaichian:2001py,Chaichian:2004za,Aschieri:2005yw,Aschieri:2005zs,qhe1,qhe2}.


\bigskip

It is known in general that various noncommutative models, including models of noncommutative field theory and noncommutative gravity allow for a representation in terms of classical commutative fields either in a compact and closed form or in a form of a perturbation expansion up to a certain order in a deformation parameter. The most known example of this kind is obtained by means of the  Seiberg-Witten (SW) map \cite{SW}, which is a field transformation that allows to rewrite a gauge theory on noncommutative space as a gauge theory on commutative space. In an attempt to map a noncommutative gauge field theory to its commutative counterpart from which it emerged as a result of deforming  an associated Poisson structure, the SW map undoubtedly  becomes  highly important. 

Related to this, it is noteworthy to recall that the SW map  helped to resolve some of the issues that had appeared by the introduction of the star-product into the action, among them being the advent of the field operator ordering ambiguities, as well as the breaking of the ordinary gauge invariance  and the problem with the charge quantisation. This map also  ensures that by going from a set of degrees of freedom describing  noncommutative gauge symmetry to a corresponding set describing  local commutative gauge symmetry the number of degrees of freedom stays the same.
This way  a number of NC deformed QFT’s could be properly defined
for arbitrary gauge group representations, which has facilitated the building of the whole range of semi-realistic NC deformed particle physics \cite{okawa,barnich2, Buric:2004ms,Buric:2006wm,Latas:2007eu,martin,tramp1,Dimitrijevic:2014dxa,DimitrijevicCiric:2018xaw} and gravity models 
\cite{Aschieri:2009ky,Aschieri:2011ng,Aschieri:2012in,Aschieri:2014xka,DimitrijevicCiric:2016qio}.

Another interesting situation where the NC theory allows for an  interpretation in terms of commutative degrees of freedom involves a class of noncommutative models which have been shown to exhibit a specific type of duality relations that are implicitly built-in  at the level of equations of motion. This  feature has been observed in some hybrid or semi-hybrid noncommutative models  investigated within a so called realization framework \cite{realization1, realization2, realization3, realization4, realization5, realization6, realization7}, which utilizes the representation of NC coordinates and NC field operators in terms of formal power series of generators of the undeformed Heisenberg algebra. The Lagrangian density in these models, which is expressed partially or completely  in terms of NC degrees of freedom (that’s why these models are being labelled as hybrid or semi-hybrid), allowed not only for a reinterpretation of the Lagrangian density in terms of commutative degrees of freedom, but also allowed for a radical refashioning of the initial  semi-hybrid NC model so that it could take on a form of
an effective commutative model  realized within a similar or even the same physical setting, but with modified system parameters. Such reinterpretation was then able to give noncommutativity a definite physical meaning. For example, in \cite{dualitymodel1,dualitymodel2,dualitymodel3,dualitymodel4,Gupta:2019cmo} it has been found that the semi-hybrid model of NC massless scalar field coupled to a classical nonrotational BTZ geometry is dual to the model of   massive commutative scalar field  probing  the  geometry of a rotating BTZ black hole. In this way, the noncommutativity took on the role of an agent medium that has  put a black hole into a state of rotation. Besides, the noncommutativity was shown to be responsible for a mass generating mechanism, as applied to a scalar probe, and for inducing certain back-reaction effects.

A similar situation has been encountered in  \cite{tureanu}, where the analogy between  NC version of the Schwarzschild black hole and the commutative Reissner–Nordstr\" om black hole with a stretched horizon was drawn. Likewise, in \cite{Trampetic:2021awu} the authors have shown  that  the minimal U(1) NCQED based on a reversible Seiberg-Witten (SW) map is equivalent to the Moyal NCQED without SW map, as manifested at the level of tree-level scattering amplitudes \cite{Trampetic:2021awu}.
In this case the equivalence between two models comes as a result of a mutual cancellation between  terms induced by the reversible SW map, which might also be viewed as being due to a presence of the specific duality relations that are inherent to the noncommutative model being considered. Similarly, the features of this kind   can be found in supersymmetric noncommutative field theories related by the theta-exact Seiberg-Witten map \cite{Horvat:2016lbd}.
In most of these cases (certainly for the case studied in \cite{dualitymodel1,dualitymodel2,dualitymodel3,dualitymodel4,Gupta:2019cmo}) the notion of duality  refers to an exact mathematical correspondence that
may be drawn  between two different physical systems having different system parameters, though governed by the same Lagrangian density and the associated equation of motion.
A duality understood that way gives  an example of the equivalence that can be established between  noncommutative and  commutative  model, where  each of these models separately describes its own respective physical system. As these two  systems that commutative and noncommutative models refer to are   actually   being
governed by the same equations of motion and the same Lagrangian density, they may too be characterized as being 
equivalent with each other.  
Therefore, referring  in the current context to a duality itself  that exists between two different physical systems, it shouldn't come as a surprise that it gets manifested
through a set of  exact mathematical transformations that connect the parameters of  these two physical systems, thus  making them dual with each other.

\bigskip

In this paper we set out to find another example of this kind
 within the semi-hybrid NC model studied in \cite{Ciric:2017rnf,DimitrijevicCiric:2019hqq,Ciric:2019urb}.
In particular, here we show that  noncommutative $U(1)_\star $ gauge theory coupled to  NC scalar field and to a classical geometry of the Reissner–Nordstr\" om type is equivalent  within a first order of deformation   to a commutative $ U(1)$
gauge theory coupled to a  commutative scalar field and to a classical geometry background, which however does not coincide with the initial RN metric, but instead represents  an effective metric which encodes the impacts of the spacetime deformation.
In other words the former model can be recast into a latter by redefining the components of the initial RN metric. In this way we end up with an effective, but equivalent description in which the redefined metric has an important role, forming one of the crucial
 building blocks that  characterize the dual system and the duality transformation itself. From now on we refer to this redefined metric as  the  effective metric characterizing the first order dual picture, or the first order effective   dual  metric in short. 
It can be viewed as having ensued from a specific deformation of the RN metric which brings in a nonvanishing $r-\phi$ component. This whole  scheme thus establishes a   first order   duality relation between two models, one (semiclassical hybrid) noncommutative and the other fully commutative,  and  two physical systems they describe.

As a further step, we study a dynamics of fermions on a curved background with a deformed spacetime structure, where as an exemplar for the curved geometry we use the RN black hole background. This study has been carried out within two different approaches: effective and formal.
The effective  approach uses standard notions of commutative differential geometry, with parallel implementation of the noncommutative-born effects through the utilisation of the type of duality just explained, where the  effective dual metrick mimicks the impacts of noncommutativity. The formal approach is a level beyond more rigorous in a sense that it attempts to stay in line and be  compatible with the requirements that  the NC $U(1)_\star $ Dirac action on a curved background remains invariant under NC gauge transformations, as well as under the undeformed local $SO(1,3)$. With that in mind, we put forth a proposal for the  NC $U(1)_\star $ Dirac action that could meet these requirements.  
As this NC $U(1)_\star $ Dirac action that we  propose is being  expressed in terms of the  NC spin $\frac{1}{2}$ field and the NC $ U(1)_\star $ gauge field that are both coupled to gravitational degrees of freedom, it is important to  stress that it will remain  invariant under  the NC  $U(1)_\star $ gauge transformations as long as  we assume  that  the gravity is unaffected  by them, i.e. 
$\delta_\star g_{\mu\nu} =0 ~ ($  or alternatively  $ ~\delta_\star e^a_{~~\mu} =   \delta_\star {\omega}_{\mu}^{~~ab} = 0 ) $.
In other words, our formal approach  assumes that the NC spin $\frac{1}{2}$ field, and the NC  $U(1)$  gauge field are the only degrees of freedom 
that get affected by the NC gauge transformations. 
In addition, we will assume that the local $SO(1,3)$ symmetry is unaffected by NC deformation.
The NC $U(1)_\star $ Dirac action that we propose is in a line with the proposals made in \cite{Aschieri:2009ky},\cite{Aschieri:2011ng}.

The main result of this paper is to show that these two different approaches surprisingly give rise to  the same equation of motion describing a dynamics of fermions on a curved space in presence of a 
NC  structure of spacetime, therefore explicitly demonstrating the equivalence between the effective and formal, more rigorous approach. Namely, what has been shown in the present paper  is that the equation of motion obtained by using the SW map for spin $\frac{1}{2}$ and gauge fields, and by varying the   NC $U(1)_\star $ Dirac action that is invariant under NC gauge transformations and $S0(1,3)_\star $ group at the end of the day appears to be the same as the equation of motion obtained by simply writing the Dirac equation on a geometric background described by the  effective dual metric.

\bigskip

Dirac equation in the context of noncommutative spaces is important from many reasons and was studied intensively in the literature.  The range of topics where it finds  application is vast, going from the high energy physics,  all  through the gravitational physics and cosmology and all the way down to the problems in condensed matter. 
In particular,  the problems related to high energy physics involve a study of  the hydrogen atom spectrum on Moyal \cite{Adorno:2009yu} and kappa-Minkowski space \cite{Harikumar:2009wv},
the impact of quantum deformation on the spin-$\frac{1}{2}$ Aharonov–Bohm effect \cite{andradesilva},
the problem of Yukawa couplings and seesaw neutrino masses in noncommutative gauge theory \cite{tramp1}, photon-neutrino interaction in  noncommutative field theory \cite{tramp3},
renormalizability and dispersion of chiral fermions in NCQED \cite{Buric:2010wd} and  the impact on neutrino oscillations due to noncommutativity of spacetime \cite{Ching:2016nfv} to name just a few.
Besides canonical and kappa-Minkowski type of noncommutativity (for a recent review see \cite{kappa-review}), other types of spacetime noncommutativity that have been frequently studied in the past include that of Snyder type (for the review see \cite{mignemi1,mignemi2}), as well as that which is usually referred to as the “spin noncommutativity”, first introduced in \cite{falomir1,falomir2}, and which could be theoretically understood as
a non-relativistic analog of the original Snyder’s model \cite{snyder}.
In \cite{spinnc}, the spin noncommutativity was obtained by means of a
consistent deformation of the Berezin-Marinov pseudoclassical model for the spinning particle
\cite{berezin}. Like the Snyder model,
spin noncommutativity exhibits preservation of the Lorentz symmetry. Within this framework a modification of the Dirac equation was proposed and a dynamics of a Dirac fermion in the presence of spin noncommutativity was studied in \cite{spinnc,Ferrari:2012bv}.

 In condensed matter, the topic of special interest is the integer and in particular the fractional quantum Hall effect and a related attempt to procure the explanation  for the latter in terms of the Dirac oscillator \cite{diracoscillator}  
 and especially in terms of the Dirac oscillator on NC space \cite{Mandal:2009we,Mandal:2012wp,quesne-tkachuk1,quesne-tkachuk2,Stetsko:2013xsa,Stetsko:2018lmt,Andrade:2013oza,Andrade:2014kxa,merad}. Due to the same reasons the study of relativistic Landau levels or Dirac-Landau levels, including the breaking of their degeneracy, becomes increasingly more important and even more so as these results were being applied to the graphene and the related nanostructures \cite{Hamil:2019knw},\cite{Oliveira:2022lhm}.

Another intriguing field where the Dirac equation under conditions of discretized spacetime was being analyzed involves deformed relativistic wave equations, namely the Klein–Gordon and
Dirac equations in a Doubly Special Relativity (DSR) scenario \cite{dsramelino,dsrmaguejo,Mignemi:2003nbf}. Besides the algebraic approach to this problem, which originally started by considering the standard real
form of the quantum anti-de Sitter algebra, $SO_q(3, 2) $ and then by consistently modifying the related coproduct \cite{tarlini}, there recently appeared  a geometric approach \cite{Relancio:2020zok,Relancio:2020rys,Pfeifer:2021tas,Franchino-Vinas:2022fkh}
to the same problem,
which is based on the geometry of a curved momentum space \cite{Amelino-Camelia:2011lvm,Amelino-Camelia:2011uwb,Gubitosi:2011hgc,Carmona:2012un,Carmona:2016obd,Carmona:2019fwf,Ivetic:2016qtz}.
While  Dirac equation obtained in \cite{tarlini} is invariant under the spin-half representation of the $ \kappa $-Poincar\'{e} algebra, it  doesn't  yield the Casimir after squaring.
 Instead, its square gives rise to the  $ \kappa $-deformed Pauli-Lubanski vector. Contrary to that,  Dirac equation obtained in \cite{Franchino-Vinas:2022fkh} gives rise to the  $ \kappa $-Poincar\'{e} Casimir upon squaring, along with having the required symmetry properties.
This geometric approach should be seen as complementary to the more spread algebraic one 
\cite{Agostini:2002yd,Aloisio:2004yz,Gosselin:2007bw,Belhadi:2011mj,bibikov,kosinski1,kosinski2}.
Finally, it is worthy to mention that  Dirac equation
 has an important role in studying  wide range of physical processes that occur near the black hole horizon, such as the scattering and absorption processes for Dirac particles \cite{scatabscrsec}, the spectral power emission of Dirac fermions \cite{page},\cite{specpow},
including the study of superradiance  \cite{chlee} and the quasinormal mode (QNM) spectrum for the fermionic 
perturbations. In the latter case, of special interest is the study of
 the impact of NC spacetime deformation  on QNM spectrum of the fermionic 
perturbations of black holes \cite{dualitymodel4}.



\bigskip

The paper is organized as follows. After a very brief review of NC deformation that we analyze  in this paper, in Section III we study semiclassical   NC $U(1)_\star $ gauge theory coupled with NC spin $\frac{1}{2}$ field and NC gravitational degrees of freedom,  whose action is invariant uder NC gauge transformations and the undeformed local  $SO(1,3) $  group.  Upon utilising the Seiberg-Witten map in order to write  NC spinor  and NC gauge fields in terms of their corresponding commutative degrees of freedom, the equation of motion for the fermion field is obtained by varying the action  over $\bar{\Psi} $. 
The effective dual metric (derived in Appendix A)  is then used in Section IV to  write a noncommutative version of the equation of motion for the fermions in a curved background of RN type, thus putting forth 
 an effective approach to the same problem that is treated in a formal, more rigorous way  in Section III. Here we find that surprisingly, both of these two approaches, formal and  effective,  yield the same final result.
 In Section V  we show that the resulting equation of motion is separable, yielding two pairs of equations, one for the angular part and the other for the radial part. Noncommutative deformation appears to affect only the radial part, with the angular part being solved by the same spin $\frac{1}{2}$ spherical harmonics as in the case of Dirac equation for the hydrogen atom in flat or Schwarzschild case. At the end, we utilise the general properties of the fermionic solutions deduced here to investigate the problem of their stability.  In particular, the issue of superradiance is considered as related to the  solutions of the Dirac equation in RN spacetime, and especially the impact of noncommutative deformation on the effect of superradiance is addressed. We end up with two Appendices. In Appendix A we demonstrate the equivalence between semiclassical NC  $U(1)_\star $ gauge theory with NC scalar field on a classical RN background and  commutative $U(1)$ gauge theory with ordinary scalar field on the background with the effective dual metric. Furthermore, we find the explicit form of the effective dual metric in a first order of deformation.  In Appendix B  we briefly discuss possible generalizations of our model   to include settings with
more nontrivial geometric backgrounds, as well as types of deformation.

\section{Preliminary settings}

A solution to Einstein equations  representing a charged non-rotating black hole  with  mass $M$  and charge $Q$  is given by the  Reissner–Nordstr\" om (RN) metric
\begin{equation}
{\rm d}s^2 = \Big(1-\frac{2MG}{r}+\frac{Q^2G}{r^2} \Big) {\rm d}t^2 - \frac{{\rm
d}r^2}{1-\frac{2MG}{r}+\frac{Q^2G}{r^2}} - r^2({\rm d}\theta^2 + \sin^2\theta{\rm d}\phi^2)
. \label{dsRN}
\end{equation}
 Being static and spherically symmetric, the  spacetime of RN black hole
 has four Killing vectors, among which  $\partial_t$ and $\partial_\phi $ are included,  and $t$ and $\phi$ are the time and polar variables  
   of the spherical  coordinate system $x^{\mu} = (t,r,\theta,\phi)$. 

In the previous paper \cite{Ciric:2017rnf} we have introduced a semiclassical model describing  a   charged NC scalar field $\hat{\Phi}$ and  NC $U(1)$ gauge field  $\hat{A}$ on a classical gravitational background of RN type. By semiclassical we mean that
while the gravitational field in this model  was assumed to be a classical degree of freedom (i.e. not deformed by noncommutativity),  the scalar  and  gauge field
propagating in that classical gravitational background were assumed to be affected by  noncommutative nature of spacetime. In a sense, we are therefore dealing with a situation where the scalar and gauge field are quantized and gravitational field is not. It is however important to stress that the gauge and scalar field are  not quantized in a sense of quantum field theory. 

The model was built by using deformation quantization techniques based on Drinfeld twist operator and the explicit twist operator that was used in the construction
was the so called angular twist operator \cite{Ciric:2017rnf, DimitrijevicCiric:2019hqq}
\begin{eqnarray}  \label{AngTwist0Phi}
\mathcal{F} &=& e^{-\frac{i}{2}\theta ^{\alpha\beta}\partial_\alpha\otimes \partial_\beta} \nonumber\\
&=& e^{-\frac{ia}{2} (\partial_t\otimes\partial_\phi - \partial_\phi\otimes\partial_t)}, 
\end{eqnarray}
with $\alpha,\beta = t,r, \theta, \phi$ and $\theta^{t\phi}= -\theta^{\phi t}=a$  as the only non-zero components of the deformation tensor $\theta^{\alpha \beta}$.
The small constant parameter $a$ is the deformation parameter  that sets up the NC scale, commonly related to the Planck length. 
This twist operator is a Killing twist, since it is built from the vector fields that are actually Killing vectors for the metric (\ref{dsRN}).
In this way it is ensured that  the geometry (\ref{dsRN})  stays unaffected by the deformation as the twist
(\ref{AngTwist0Phi}) does not act on the RN metric.

The star product, the wedge star product between forms, the coproduct and other structural maps of the related symmetry algebra can all be obtained from  the twist operator
(\ref{AngTwist0Phi}). In particular, the star product between functions is given by
\begin{eqnarray}  \label{fStarg0Phi}
f\star g &=&  \mu \circ {\mathcal{F}}^{-1} \left( f\otimes g  \right)   = \mu \{ e^{\frac{ia}{2} (\partial_t \otimes \partial_\phi - \partial_\phi \otimes
\partial_t)}
f\otimes g \}\nonumber\\
&=& fg + \frac{ia}{2}
(\partial_t f(\partial_\phi g) - \partial_t g(\partial_\phi f)) + 
\mathcal{O}(a^2) ,
\end{eqnarray}
where the map $\mu$  represents the usual pointwise multiplication.
The remaining ingredients of the differential calculus are described in \cite{Ciric:2017rnf}.

\section{Spinor field on the noncommutative RN background: formal analysis}

A massive and charged spinor field  $\Psi $ on a fixed gravitational background can be described by the following action
\begin{equation}
S = \int \d ^4x \;  |e|   \bar{\Psi} \Big( i\gamma^\mu D_\mu \Psi -m\Psi\Big). \label{SPsiNula}
\end{equation}
The mass and the charge of the spinor field $\Psi$ are respectively $m$ and $q$. The determinant of the vierbein $e^a_{~~\mu}$ we label with $|e|=\sqrt{-g}$ and the covariant derivative $D_\mu$ includes both the spin connection $\omega_\mu$ and the $U(1)$ gauge field $A_\mu$
\begin{equation}
D_\mu\Psi = \partial_{\mu} \Psi - \frac{i}{2}  \omega_{\mu}^{~~ab} \Sigma_{ab}\Psi -iqA_\mu\Psi .\label{CovDerivativeNula} 
\end{equation}
Matrices $\Sigma^{ab}$ are the (herimtean) generators of the local Lorentz transforamtions and they close the Lorentz algebra
\begin{equation}
[\Sigma^{ab} , \Sigma^{cd}] = i(\eta^{ad}\Sigma^{bc} + \eta^{bc}\Sigma^{ad} - \eta^{ac}\Sigma^{bd} - \eta^{bd}\Sigma^{ac}) .\nn
\end{equation}
The spin connection is not an independent field, but a function of $e^a_{~~\mu}$, calculated from the torsion free condition
\begin{equation}
T^a_{\mu\nu} = \nabla_\mu e^a_{~~\nu} - \nabla_\nu e^a_{~~\mu} = 0,\nn  
\end{equation}
with $\nabla_\mu e^a_{~~\nu} = \partial_\mu e^a_{~~\nu} +  \omega_{\mu ~~~ b}^{~~a} ~   e^b_{~~\nu}$. For more details on the spin connection, vierbeins and the notation we use, we refer to the beginning of Section IV.

The action (\ref{SPsiNula}) is invariant under the local $U(1)$ transformations
\begin{equation}
\delta_\alpha \Psi = i\alpha (x)\Psi, \quad \delta_\alpha \bar{\Psi} = -i\bar{\Psi}\alpha (x), \quad \delta_\alpha A_\mu = \frac{1}{q}\partial_\mu \alpha (x).\label{U1Komutativne} 
\end{equation}
Note that these transformations do not act on the gravitational background, that is on $e^a_{~~\mu}$ and $\omega_\mu$. The action (\ref{SPsiNula}) is also invariant under the  
general coordinate transformations
and the local $SO(1,3)$ symmetry. In this paper we use the semiclassical analysis and promote only the $U(1)$ gauge symmetry to the noncommutative $U(1)_\star$ gauge symmetry. In our future work we will lift this approximation and allow for the noncommutative local $SO(1,3)_\star$ symmetry.
The equation of motion for the spinor field $\Psi$ is obtained by varying the action (\ref{SPsiNula}) with respect to $\bar{\Psi}$ and it is given by
\begin{equation}
i\gamma^\mu \Big(\partial_\mu \Psi -i \omega_\mu\Psi - iqA_\mu\Psi\Big) -m\Psi =0 .\label{EoMPsiNula}
\end{equation}

Following the steps from \cite{Ciric:2017rnf}, we now introduce an action functional that describes the NC $U(1)_\star$ gauge theory of a charged spinor field on the RN background\footnote{Actually, it can be any  background with Killing vectors
$\partial_t$ and $\partial_\phi$.} (\ref{dsRN}) 
\begin{equation}    
S_\star = \int \d ^4x ~ |e| \star\bar{\hat{\Psi}} \star\Big( i\gamma^\mu \big( \partial_{\mu} \hat{\Psi} - i\omega_\mu \star \hat{\Psi} -iq\hat{A_\mu}\star \hat{\Psi} \big) -m\hat{\Psi}\Big). \label{NCActionPsi}
\end{equation}
Noncommutative fields are now labeled with a $\hat{\>}$ and the $\star$-product is given by (\ref{fStarg0Phi}).  
One can show that this action is invariant under the following infinitesimal $U(1)_\star$ gauge transformations:
\begin{eqnarray}
\delta_\star \hat{\Psi} &=& i\hat{\Lambda} \star \hat{\Psi}, \nonumber  \\
\delta_\star \hat{A}_\mu &=& \partial_\mu\hat{\Lambda} + i  \big( \hat{\Lambda} \star \hat{A}_\mu - \hat{A}_\mu \star \hat{\Lambda}  \big),
\label{NCGaugeTransf}\\
\delta_\star \omega_\mu &=& \delta_\star e^a_{~~\mu} =0, \nonumber
\end{eqnarray}
where $\hat{\Lambda}$ is the NC gauge parameter. In particular, note that
\begin{equation}
\delta_\star D_\mu \hat{\Psi} = i\hat{\Lambda} \star D_\mu \hat{\Psi} \nn
\end{equation}
since the twist  (\ref{AngTwist0Phi}) does not act on the gravitational field and therefore  $\omega_\mu\star\Lambda = \omega_\mu \cdot \Lambda = \Lambda \star \omega_\mu$. Note that the action (\ref{NCActionPsi}) can be written in a more geometric way \cite{PaoloLeonardo} such that the  general coordinate transformation  invariance is manifest
\begin{equation}
S \sim \int \Big( (\overline{D{\hat{\Psi}}})_B \star \hat{\Psi}_A -
\bar{\hat{\Psi}}_B \star (D\hat{\Psi})_A \Big)\wedge_\star (e\wedge_\star                                            
e\wedge_\star e \gamma_5)_{BA}, \label{NCSpinorGeom}
\end{equation}
with spinor indices $A, B$ explicitly written and the vierbein one form $e$. The covariant derivative  one-form  is given by $D\hat{\Psi} = {\rm d}\hat{\Psi} -i\omega \star  \hat{\Psi} - i\hat{A}  \star  \hat{\Psi}$, with the gauge potential one-form $\hat{A}$ and the spin connection one-form $\omega$. When the Killing twist is used and the action (\ref{NCSpinorGeom}) is expanded in a chosen coordinate basis, the action (\ref{NCActionPsi}) is obtained.

Similarly to \cite{Ciric:2017rnf}, we can add the action for the NC $U(1)_\star$ gauge field $\hat{A}_\mu$ to (\ref{NCActionPsi}), promoting the gauge field $\hat{A}_\mu$ into a dynamical field. However, since later on we will be interested in propagation of NC spinor field on the fixed RN background, we do not write the action for the NC $U(1)_\star$ gauge field $\hat{A}_\mu$ explicitly here.

To simplify the calcualtion, from now on we redefine $A_\mu =
qA_\mu$. Then we use the  Seiberg-Witten (SW)-map \cite{SW},\cite{jssw} in order  to express  NC fields $ \hat{\Psi}$ and $\hat{A}_\mu$ as functions of the corresponding commutative fields
and the deformation parameter $a$. The SW-map assumes an expansion  in orders of the deformation parameter  and this expansion is known to all orders for an arbitrary Abelian twist deformation \cite{jssw},\cite{Aschieri:2011ng},\cite{Aschieri:2012in}  of which the twist (\ref{AngTwist0Phi})  is only one example.
For  the twist operator (\ref{AngTwist0Phi}), SW-map gives rise to the following expansions for the fields:
\begin{eqnarray}
\hat{\Psi} &=& \Psi -\frac{1}{2}\theta^{\rho\sigma}A_\rho(\partial_\sigma\Psi) , \label{HatPsi}\\
\hat{A}_\mu &=& A_\mu -\frac{1}{2}\theta^{\rho\sigma}A_\rho(\partial_\sigma A_{\mu} +
F_{\sigma\mu}).  \label{HatA}
\end{eqnarray}
The expanded action  up to first order in $a$  is given by
\begin{eqnarray}
S_\star &=& \int \d ^4x ~ |e| \bar{\Psi} \Big( i\gamma^\mu D_\mu \Psi -m\Psi\Big) \nn\\
&& +\frac{1}{2}\theta^{\alpha\beta}\Big( -iF_{\mu\alpha}\bar{\Psi}\gamma^\mu D^{\mbox{\tiny{U(1)}}}_\beta\Psi
-\frac{i}{2}\bar{\Psi}\gamma^\mu \omega_\mu F_{\alpha\beta}\Psi 
-\frac{1}{2}F_{\alpha\beta}\bar{\Psi} \big(i\gamma^\mu D^{\mbox{\tiny{U(1)}}}_\mu \Psi -m\Psi\big) .\nn
\end{eqnarray}
Remebering that $F_{\alpha\beta} = \partial_\alpha A_\beta - \partial_\beta A_\alpha$ and
choosing the electromagnetic potential to be that of the RN black hole,
the only non-zero component of $F_{\alpha\beta}$ is $F_{rt} = \frac{qQ}{r^2}$. This leads to a simplified NC action
\begin{equation}
S_\star = \int \d ^4x ~ |e| \bar{\Psi} \Big( i\gamma^\mu D_\mu \Psi -m\Psi\Big) -\frac{i}{2}\theta^{\alpha\beta}\bar{\Psi} F_{\mu\alpha}\gamma^\mu D^{\mbox{\tiny{U(1)}}}_\beta\Psi \label{SNCExpanded}
\end{equation}
and the corresponding equation of motion for the spinor $\Psi$
\begin{equation}
i\gamma^\mu \Big(\partial_\mu \Psi -i \omega_\mu\Psi - iA_\mu\Psi\Big) -m\Psi - \frac{ia}{2}F_{rt}\gamma^r\partial_\phi \Psi = 0 .
\end{equation}
Inserting the explicit expressions for $F_{rt}$ and $\gamma^{r} = e_a^{~~r} \gamma^{a},$ this equation reduces to
\begin{equation}     \label{EoMPsiNC} 
   i\gamma^\mu \Big(\partial_\mu \Psi -i \omega_\mu\Psi - iA_\mu\Psi\Big) -m\Psi -
    \frac{ia}{2}  \frac{qQ}{r^2} \sqrt{f} \gamma^1 \partial_{\phi} \Psi = 0 .
\end{equation}

For a later  comparison with the result that will be obtained in an effective approach, it is instructive to write this equation in terms of two-component spinors
 $\Psi =  \left(\begin{matrix} \Psi_1  \\ \Psi_2  \\   \end{matrix}\right)$.
Note that the spin connection part $\omega_{\mu}$ and vielbeins refer to RN background{\footnote{ How they look like in  RN background may be inferred from
relations (\ref{spinpart}),
 (\ref{metrictetrad}) and  (\ref{metrictetradInverse}), by taking into account that $\omega_\mu \equiv -\frac{1}{2} \omega_{\mu}^{~~cd} \Sigma_{cd}$  and by setting $a$ to zero. }.
\begin{equation} \label{easiercomparison}
\begin{split}
&  i\frac{1}{\sqrt{f}} i \left( \begin{array}{ccccc}
  0  & \mathbb{1}  \\
   \mathbb{1}   & 0  \\ 
\end{array} \right)
 \partial_t   \Psi +  i \sqrt{f} i \left( \begin{array}{ccccc}
  0  & \sigma_3  \\
   -\sigma_3   & 0  \\ 
\end{array} \right)  \partial_r \Psi +  i\frac{1}{r}  i \left( \begin{array}{ccccc}
  0  & \sigma_1  \\
   -\sigma_1   & 0  \\ 
\end{array} \right) \partial_\theta \Psi  + i  \frac{1}{r\sin \theta} i \left( \begin{array}{ccccc}
  0  & \sigma_2  \\
   -\sigma_2   & 0  \\ 
\end{array} \right) \partial_\phi \Psi   \\
& \quad +   \Big(  e_0^{~~t}  \gamma^0 \omega_t + e_2^{~~\theta}  \gamma^2 \omega_\theta +
    e_3^{~~\phi}  \gamma^3 \omega_\phi  \Big) \Psi +  e_0^{~~t}  \gamma^0 A_t \Psi - m\Psi + \frac{a}{2} \frac{qQ}{r^2} \sqrt{f}
  \left( \begin{array}{ccccc}
  0  & \sigma_3  \\
   -\sigma_3   & 0  \\ 
\end{array} \right) \partial_\phi \Psi = 0.
\end{split}
\end{equation} 

The particular model of NC gauge theory  applied to NC spinor field, that we consider here,  involves  NC spinor field which  is minimally coupled to 
both,  NC $U(1)$  gauge field and  classical gravitational field of the RN background. While the gauge field itself is fixed to be the Coulomb field, but only until  after rejecting all except the first order terms in the SW expansion and varying the action to get the equation of motion, the gravitational field is fixed from the very beginning to be that of the RN background. In practice this means that the only propagating degrees of freedom in this model are those of the matter fields. We point out that the working setting just described is completely analogous (even identical) to the one that we used in our previous work \cite{Ciric:2017rnf}  when studying a particular model of NC gauge theory, as applied to NC scalar field.
The main assumption of this setting is that a gravitational field is being considered as a  classical (commutative) object, and matter fields along with a gauge field are being considered as noncommutative 
objects. From this reason
 we term this kind of working framework as {\it{semiclassical}}, bearing on the fact that it does not correspond to a full NC gauge theory, but only to a description of  NC matter field in  a particular setup (static charge, black hole geometry).
The immediate consequence of the  gravitational degrees of freedom (either components of the metric or vielbeins) being classical is that they do not change/transform
under  infinitesimal NC gauge transformations. Therefore, an immutability of the gravitational degrees of freedom is an assumption and a starting point of this particular NC framework, and not its consequence.

Mathematically, the semiclassical approximation manifests itself in (\ref{SNCExpanded}) in the following way: the covariant derivative $D_\mu \Psi = \partial_\mu\Psi - iA_\mu \Psi - i\omega_\mu \Psi$ includes both the electromagnetic  $(U(1)) $  and the gravitational part, while the covariant derivative $D^{\mbox{\tiny{U(1)}}}_\beta\Psi = \partial_\beta\Psi - iA_\beta \Psi$ has only the electromagnetic part. In the NC correction only the $U(1)$  part appears.

\section{Spinor fields on the noncommutative RN background: effective approach}
In our previous paper \cite{Ciric:2017rnf} we have analyzed a propagation of the NC scalar field in the RN background. The equation of motion governing the evolution of the scalar field is given by

\begin{equation}  \label{extendedKG}
\Big( \frac{1}{f}\partial^2_t -\Delta + (1-f)\partial_r^2 
+\frac{2MG}{r^2}\partial_r + 2iqQ\frac{1}{rf}\partial_t -\frac{q^2Q^2}{r^2f} -\mu^2 \Big) \Phi  +\frac{aqQ}{r^3}
\Big( (\frac{MG}{r}-\frac{GQ^2}{r^2})\partial_\phi
+ rf\partial_r\partial_\phi \Big) \Phi =0 ,  
\end{equation}
with $f = 1-\frac{2MG}{r}+\frac{Q^2G}{r^2}$. In \cite{dualitymodel2,dualitymodel3}, within first order of deformation, an equivalence between the NC scalar field on the non-rotating BTZ background and a
commutative scalar field on the rotating BTZ background was established. We will follow that idea here and try to undestand if there is an effective description of the NC scalar field on the RN background.


It can be shown that the equation of motion  (\ref{extendedKG})  
may be rewritten as the equation of motion governing  a charged commutative scalar field with the same charge $q$ as its NC counterpart, and propagating in some effective metric. That this process of finding a metric from the given equation of motion
 can indeed be carried out\footnote{It can be carried out at least for the type of geometry and deformation considered in this paper, i.e. a deformation with the Killing twist operator. For more general examples of deformation, see Appendix B.} within a first order of deformation is shown in Appendix A. There
the first order effective dual metric has been derived and shown to pick up the form of 
 a modified RN geometry
\begin{equation} \label{NCdsRN}   
  {\rm d}s^2 = \Big(1-\frac{2MG}{r}+\frac{Q^2G}{r^2} \Big) {\rm d}t^2 - \frac{{\rm
d}r^2}{1-\frac{2MG}{r}+\frac{Q^2G}{r^2}} - aqQ \sin^2 \theta {\rm d} r {\rm d} \phi - r^2({\rm d}\theta^2 + \sin^2\theta{\rm d}\phi^2 ).
\end{equation}
It appears that  new, first order effective dual  metric (\ref{NCdsRN})  acquires an additional off-diagonal term which is induced purely by noncommutative nature of spacetime. This  feature  comes into play only in the presence of charged matter. Unlike in the case of scalar field in the (NC) BTZ background, in this case the effective metric cannot be interpreted as a metric of a background rotating geometry (of either RN or any other type).


Having established the  effective metric (\ref{NCdsRN}), we now investigate the propagation of a charged massive spinor field $\Psi$  in this geometry. In particular, it is interesting to see if this effective approach agrees with the more rigorous approach from Section II. The Dirac equation in a curved  background given by the effective noncommutetive (NC)  metric
\begin{equation}\label{diracnovo}
( i \gamma^a \nabla_a - m)\Psi=0,
\end{equation}
where the Latin indices such as $a, ~(a =0,1,2,3)$ refer to intrinsic coordinates and $\gamma^a$ are the standard flat space Dirac gamma matrices, $~\{ \gamma_a, \gamma_b \} = 2 \eta_{ab},~$ where
\begin{equation} \label{metriceta}
  \eta_{ab} = \eta^{ab} =
\left( \begin{array}{ccccc}
  +1  & 0  & 0 & 0  \\
   0   & -1 & 0 & 0  \\ 
   0  & 0 & -1 &  0  \\
   0  & 0 & 0 & -1  \\
\end{array} \right).  
\end{equation}
If in addition, the spinor field is charged, this gives rise to a Dirac equation in which the gauge potential $A_{\mu}$  is minimally coupled to a Dirac operator on a curved background
\begin{equation}\label{diracnovogauge}
\bigg( i \gamma^a ( \nabla_a -i A_a ) - m \bigg)\Psi= \bigg( i \gamma^a e_a^{~~\mu} ( \nabla_{\mu} -i A_{\mu} ) - m \bigg)\Psi =0.
\end{equation}
The gravitational covariant derivative $\nabla_{\mu}$ is defined as  $\nabla_{\mu} \Psi = \partial_{\mu} \Psi  - \frac{i}{2} \omega_{\mu}^{~~ab} \Sigma_{ab} \Psi.$
The Dirac operator $\gamma^a \nabla_a$ on a curved space
is introduced in terms of tetrads (vierbeins) $~ e^a_{~~\mu} ~$ and their inverse  $~ e_a^{~~\mu}, ~$ satisfying
$~ e^a_{~~\mu}  e_a^{~~\nu} = \delta_{\mu}^{~~\nu} ~$ and  $~ e^a_{~~\mu}  e_b^{~~\mu} = \delta^{a}_{~~b}. ~$  Tetrads written in components are
$~  e^a_{~~\mu} = ( e^a_{~~t}, e^a_{~~r}, e^a_{~~\theta}, e^a_{~~\phi }) ~ $ and $ ~    e_a^{~~\mu} = ( e_0^{~~\mu}, e_1^{~~\mu}, e_2^{~~\mu}, e_3^{~~\mu}). ~$ 
 They also satisfy 
$~g_{\mu \nu} =  e^a_{~~\mu}  e^b_{~~\nu}  \eta_{ab} ~$ and $~g^{\mu \nu} =  e_a^{~~\mu}  e_b^{~~\nu}  \eta^{ab}.$  
In what follows we use the setting defined in \cite{Dolan:2015eua} with the vierbein frame chosen to be
\begin{equation} \label{metrictetrad}
  e^a_{~~\mu} =
\left( \begin{array}{ccccc}
  \sqrt{f} & 0  & 0 & 0  \\
   0   & \frac{1}{\sqrt{f}} & 0 & 0  \\ 
   0  & 0 & r &  0 \\
 0  & \frac{aqQ }{2r} \sin \theta & 0 &  r \sin \theta \\
\end{array} \right)
\end{equation}
with the corresponding  inverse matrix
\begin{equation}\label{metrictetradInverse}
e_a^{~~\mu} =
\left( \begin{array}{ccccc}
  \frac{1}{\sqrt{f}} & 0  & 0 & 0 \\
   0   & \sqrt{f} & 0 &  -\frac{aqQ}{2 r^2} \sqrt{f}  \\ 
   0  & 0 &  \frac{1}{r} & 0 \\
  0  & 0 & 0 &  \frac{1}{r \sin \theta }  \\
\end{array} \right) .
\end{equation}
The representation of gamma matrices is 
\begin{equation}
\begin{array}{lll}
\gamma^0 = i  \tilde{\gamma}^0 =
 i \left( \begin{array}{ccccc}
  0  & \mathbb{1}  \\
   \mathbb{1}   & 0  \\ 
\end{array} \right),  & & 
 \gamma^1 =   i \tilde{\gamma}^3 =
i \left( \begin{array}{ccccc}
  0  & \sigma_3  \\
   -\sigma_3   & 0  \\  
 \end{array} \right), \\
&&\\
\gamma^2 =   i \tilde{\gamma}^1 =
i\left( \begin{array}{ccccc}
  0  & \sigma_1  \\
   -\sigma_1   & 0  \\ 
\end{array} \right),  & &
 \gamma^3 =   i \tilde{\gamma}^2 =
i\left( \begin{array}{ccccc}
  0  & \sigma_2  \\
   -\sigma_2   & 0  \\ 
\end{array} \right),
\end{array} \label{gammarep}
\end{equation}
where $\tilde{\gamma}^0$,  $\tilde{\gamma}^1$, $\tilde{\gamma}^2$ and $\tilde{\gamma}^3$ are gamma matrices in chiral (Weyl) representation, while  $~ \sigma_i, ~ (i=1,2,3)~$ are the usual Pauli matrices
\begin{equation} \label{Pauli}
   \sigma_1 =
\left( \begin{array}{ccccc}
  0  & 1  \\
   1   & 0  \\  
 \end{array} \right), \quad \quad 
 \sigma_2 =
\left( \begin{array}{ccccc}
  0  & -i  \\
   i   & 0  \\ 
\end{array} \right), \quad \quad 
  \sigma_3 =
\left( \begin{array}{ccccc}
  1  & 0  \\
   0   & -1  \\ 
\end{array} \right).  
\end{equation} 
By writing out a detailed structure of the covariant derivative $~\nabla_a, ~$  the Dirac equation (\ref{diracnovogauge}) takes the form\footnote{Since  implementation  of (\ref{gammarep}) as our representation of $\gamma-$matrices involves a flip in their hermiticity properties (hermitian turns into antihermitian), the covariant derivative gets changed,  $\nabla_{\mu} = \partial_{\mu} - \frac{i}{2} {\omega}_{\mu}^{~~cd}  \frac{i}{4} [\gamma_c, \gamma_d]    \rightarrow  \nabla_{\mu} = \partial_{\mu} - \frac{1}{2} {\omega}_{\mu}^{~~cd}  \frac{1}{4} [\gamma_c, \gamma_d] $, which amounts to changing the sign in the covariant derivative in front of the spin part, $\nabla_{\mu} =  \partial_{\mu} 
+ \frac{i}{2} {\omega}_{\mu}^{~~cd}  \Sigma_{cd}    = \partial_{\mu} 
+ \frac{i}{2} {\omega}_{\mu}^{~~cd}  \frac{i}{4} [\gamma_c, \gamma_d]$ .}

\begin{equation}\label{diracnovo1}
\bigg[ i \gamma^a e_a^{~~\mu} \bigg( \partial_{\mu} + \frac{i}{2} {\omega}_{\mu}^{~~cd} \Sigma_{cd} 
   -i A_{\mu}  \bigg) - m  \bigg]\Psi=0.
\end{equation}
Here $~\Sigma_{cd} = \frac{i}{4} [\gamma_c, \gamma_d]~$
and the coefficients of the spin connection $~\omega_{\mu}^{~~ab}~$ are given by
\begin{equation} \label{spinconnection}
\begin{split} 
  \omega_{\mu}^{~~ab} & = e^a_{~~\nu} \eta^{bc} \partial_{\mu} e_c^{~~\nu} + e^a_{~~\nu} \eta^{bc} e_c^{~~\lambda} \Gamma^{\nu}_{~~\mu \lambda} \\
  & = \frac{1}{2} e^{a\nu} \bigg( \partial_{\mu} e^b_{~~\nu} - \partial_{\nu} e^b_{~~\mu} \bigg)
 - \frac{1}{2} e^{b\nu} \bigg( \partial_{\mu} e^a_{~~\nu} - \partial_{\nu} e^a_{~~\mu} \bigg)
 - \frac{1}{2} e^{a\rho} e^{b\sigma} \bigg( \partial_{\rho} e_{c\sigma} - \partial_{\sigma} e_{c\rho} \bigg) e^c_{~~\mu},   \nonumber
\end{split}
\end{equation}
where $~\Gamma^{\nu}_{~~\mu \lambda} = \frac{1}{2} g^{\nu \delta} \bigg( \partial_{\mu} g_{\delta \lambda} + 
\partial_{\lambda} g_{\mu \delta} - \partial_{\delta} g_{\mu \lambda} \bigg)$ are the coefficients of the affine connection. Note that ${\omega}_{\mu}^{~~ab} = - {\omega}_{\mu}^{~~ba}$.

With the tetrads given in   (\ref{metrictetrad}), one gets that  the only non zero components of the spin connection  are
\begin{equation} \label{spinconnection1}
\begin{array}{lll}
\omega_{t}^{~~01}    = -\omega_{t}^{~~10} = -\frac{Mr - Q^2}{r^3}, 
& & \omega_{\theta}^{~~12}    = -\omega_{\theta}^{~~21} = \sqrt{f}, \\
&&\\
\omega_{\phi}^{~~13}    = -\omega_{\phi}^{~~31} = \sqrt{f} \sin \theta,  & &
    \omega_{\phi}^{~~23}    = -\omega_{\phi}^{~~32} =  \cos \theta, \\
&&\\
\omega_{r}^{~~23}    = -\omega_{r}^{~~32} = \frac{aqQ}{2 r^2} \cos \theta,  & &
    \omega_{r}^{~~13}    = -\omega_{r}^{~~31} =    \frac {aqQ \sqrt{f}}{2 r^2}\sin \theta. 
\end{array}
\end{equation}
In  subsequent analysis we will also use the sums $\omega_{\mu}^{~~cd} \Sigma_{cd}$:
\begin{equation} \label{spinpart}
\begin{split}
\omega_{t}^{~~cd} \Sigma_{cd} &= 2 \omega_{t}^{~~01} \Sigma_{01} =-2 \frac{Mr - Q^2}{r^3} \frac{i}{4} [\gamma_0, \gamma_1]= -i\frac{Mr - Q^2}{r^3}
 \left( \begin{array}{ccccc}
-\sigma_3   &  0 \\
0   & \sigma_3   \\  
\end{array} \right),
 \\
\omega_{r}^{~~cd} \Sigma_{cd} &= 2 \omega_{r}^{~~23} \Sigma_{23} + 2 \omega_{r}^{~~13} \Sigma_{13}
= - \frac{aqQ}{2r^2} \cos \theta  \left( \begin{array}{ccccc}
  \sigma_3   &  0 \\
   0   & \sigma_3   \\  
\end{array} \right) 
+ \frac{aqQ \sqrt{f}}{2r^2} \sin \theta  \left( \begin{array}{ccccc}
  \sigma_1   &  0 \\
   0   & \sigma_1   \\  
 \end{array} \right) ,   
\\
  \omega_{\theta}^{~~cd} \Sigma_{cd} &= 2 \omega_{\theta}^{~~12} \Sigma_{12} = - \sqrt{f}
 \left( \begin{array}{ccccc}
  \sigma_2   &  0 \\
   0   & \sigma_2   \\  
 \end{array} \right),
\\
 \omega_{\phi}^{~~cd} \Sigma_{cd} &= 2 \omega_{\phi}^{~~13} \Sigma_{13} + 2 \omega_{\phi}^{~~23} \Sigma_{23}
= \sqrt{f}  \sin \theta  \left( \begin{array}{ccccc}
  \sigma_1   &  0 \\
   0   & \sigma_1   \\  
 \end{array} \right) 
- \cos \theta  \left( \begin{array}{ccccc}
  \sigma_3   &  0 \\
   0   & \sigma_3   \\  
 \end{array} \right) ,   \\
\end{split}
\end{equation}
Inserting these into (\ref{diracnovo1}) leads  to the Dirac  equation
\begin{equation} \label{diracnovo2}
\begin{split}
& \bigg[  i \gamma^0  e_0^{~~t}  \bigg( \partial_t - iA_t  + \frac{i}{2}  \omega_{t}^{~~cd} \Sigma_{cd} \bigg) +
    i \gamma^1  e_1^{~~r}  \bigg( \partial_r  - iA_r + \frac{i}{2}  \omega_{r}^{~~cd} \Sigma_{cd} \bigg)  +
  i \gamma^2  e_2^{~~\theta}  \bigg( \partial_{\theta} - iA_\theta + \frac{i}{2}  \omega_{\theta}^{~~cd} \Sigma_{cd} \bigg)   \\
& + i \gamma^1  e_1^{~~\phi}  \bigg( \partial_{\phi} - iA_\phi + \frac{i}{2}  \omega_{\phi}^{~~cd} \Sigma_{cd} \bigg) +
     i \gamma^3  e_3^{~~\phi}  \bigg( \partial_{\phi} - iA_\phi + \frac{i}{2}  \omega_{\phi}^{~~cd} \Sigma_{cd} \bigg)  - m \bigg] \Psi
   =0. \\
\end{split}
\end{equation}
With the spinor field $\Psi$ written in terms of  two two-component spinors $\Psi_1$ and $\Psi_2$, namely $\Psi =
  \left(\begin{matrix} \Psi_1  \\ \Psi_2  \\   \end{matrix}\right)$ and the gauge potential $A_{\mu}   = (A_t, \vec{A}) = (-\frac{qQ}{r}, \vec{0}),$
the equation (\ref{diracnovo2}) splits into  two two-component equations
\begin{equation} \label{diracnovo4}
\begin{split}
& \bigg[  - \frac{1}{\sqrt{f}} \mathbb{1} \partial_t   - \sqrt{f} \sigma_3 \partial_r - \frac{1}{2}  \frac{Mr - Q^2}{r^3} \frac{1}{\sqrt{f}} \sigma_3  - \frac{\sqrt{f}}{r} \sigma_3 -\frac{1}{r} \sigma_1 \partial_{\theta}   \\
& \quad + \frac{aqQ}{2r^2} \sqrt{f} \sigma_3 \partial_{\phi}   - \frac{1}{r\sin \theta} \sigma_2 \partial_{\phi}    
- \frac{1}{2r} \cot \theta \sigma_1   - \frac{iqQ}{r \sqrt{f}} \mathbb{1}      \bigg] \Psi_2 -  m \mathbb{1}  \Psi_1 =0, \\
& \bigg[- \frac{1}{\sqrt{f}} \mathbb{1} \partial_t +   \frac{1}{2}  \frac{Mr - Q^2}{r^3} \frac{1}{\sqrt{f}} \sigma_3 +
     \sqrt{f} \sigma_3 \partial_r  + \frac{\sqrt{f}}{r} \sigma_3 + \frac{1}{r} \sigma_1 \partial_{\theta}   \\
&\quad - \frac{aqQ}{2r^2} \sqrt{f} \sigma_3 \partial_{\phi}
+ \frac{1}{r\sin \theta} \sigma_2 \partial_{\phi}  
   + \frac{1}{2r} \cot \theta \sigma_1  - \frac{iqQ}{r \sqrt{f}} \mathbb{1}   \bigg] \Psi_1 - m \mathbb{1}  \Psi_2 =0.
\end{split}
\end{equation}
We see that these equations have the same form as the equation (\ref{EoMPsiNC}). The only NC correction is of the form  $~   \frac{aqQ}{2 r^2} \sqrt{f} \gamma^1 \partial_{\phi} \Psi. $ Therefore we can conclude that the rigorous approach of the NC gauge theory and the SW expansion described in Section III and the effective approach described here lead to the same result. However, two comments are in order.
 Firstly, our results are valid up to first order in the deformation parameter $a,$ implying that  the linearized equations of motion for a spinor field in SW expansion turn out to be the same as equations of motion in a perturbed (first order effective dual) metric.
Secondly, the result in Section III was deduced using the semiclassical approximation. In our future work we plan to investigate if this duality holds more generally.

\section{Discussion and outlook}

Before we go on to analyze the set of equations (\ref{diracnovo4}), note that the main result of our paper may be restated
in a slightly different way, by using interpretation in terms of a process of reversed engineering described in Section IV and Appendix B, which has led to
the first order effective dual metric (\ref{NCdsRN}).
In this respect, it is important to emphasize that
the process of reversed engineering as applied to the equations of motion resulting from two different NC gauge theory models, one for NC scalar field, and the other for  NC spin one-half field, may not necessarily lead to the same first order effective dual metric. Quite opposite,   it would be highly unlikely for this to happen.
However, within our semiclassical model of NC gauge theory, we have manifestly demonstrated that  this is indeed the case, and this constitutes the  main result of our paper.
 More precisely, two first order effective dual metrics, obtained by two different and
independent back engineering processes, one applied on the particular NC scalar field model and the other on the related NC spin one-half 
model, are the same! We have shown this by establishing the first order equivalence between the formal and the  effective approach
to the semiclassical $U(1)$ gauge theory applied on NC spinor field. This way we have gone around and bypassed the process of reversed engineering on
the equation of motion for the NC spinor field.
To be more precise, we took one of the first order effective dual metrics, the one obtained by  back engineering 
the NC scalar field equation of motion
and then we used this metric to write down the equation of motion for the ordinary commutative spinor field.
Interestingly enough, it  turned out that this equation is the same as the equation of motion obtained in a more formal approach to the semiclassical model of NC gauge theory, as applied to NC spin one-half field.

Let us now analyze the equation (\ref{diracnovo4}) in more detail. This equation can be used to study various effects, such as NC spinor bouned states or quasinormal modes in the RN background.

In order to solve (\ref{diracnovo4}) for the wavefunction $\Psi \equiv \Psi(t, r,\theta, \phi)$, we  follow \cite{Dolan:2015eua} and take the ansatz
\begin{equation} \label{anzatz}
\Psi =
  \left(\begin{matrix} \Psi_1  \\ \Psi_2  \\   \end{matrix}\right)
  = \left(\begin{matrix} \Psi^{(1)}_1  \\ \Psi^{(2)}_1  \\  \Psi^{(1)}_2  \\ \Psi^{(2)}_2  \\ \end{matrix}\right)
  = \frac{1}{r} f^{-1/4} \left(\begin{matrix} \psi_1  \\ \psi_2  \\ \end{matrix}\right)
= {\big( r^4 f \big) }^{-1/4} \left(\begin{matrix} \psi_1  \\ \psi_2  \\ \end{matrix}\right)  =
   {\bigg( r^4  - 2Mr^3 + Q^2 r^2 \bigg) }^{-1/4} \left(\begin{matrix} \psi^{(1)}_1  \\ \psi^{(2)}_1  \\  \psi^{(1)}_2  \\ \psi^{(2)}_2  \\ \end{matrix}\right).
\end{equation}
After plugging this ansatz into (\ref{diracnovo4}) and performing some simplicifations, the set of equations (\ref{diracnovo4}) reduces to
\begin{equation} \label{diracnovo5}
\begin{split}
 \bigg[  - \frac{r}{\sqrt{f}} \mathbb{1} \partial_t   -  r\sqrt{f} \sigma_3 \partial_r  - \sigma_1 \partial_{\theta} 
- \frac{1}{\sin \theta} \sigma_2 \partial_{\phi} - \frac{1}{2} \cot \theta \sigma_1 
+ \frac{aqQ}{2r} \sqrt{f} \sigma_3 \partial_{\phi} - \frac{iqQ}{ \sqrt{f}} \mathbb{1}  \bigg] \psi_2 -  mr \mathbb{1}  \psi_1=0,
 \\
\bigg[- \frac{r}{\sqrt{f}} \mathbb{1} \partial_t +  r  \sqrt{f} \sigma_3 \partial_r  +  \sigma_1 \partial_{\theta}
+ \frac{1}{\sin \theta} \sigma_2 \partial_{\phi} + \frac{1}{2} \cot \theta \sigma_1 
- \frac{aqQ}{2r} \sqrt{f} \sigma_3 \partial_{\phi} - \frac{iqQ}{ \sqrt{f}} \mathbb{1}
\bigg] \psi_1 - m r \mathbb{1}  \psi_2 =0.
\end{split}
\end{equation}
We further make a factorization of the spinor wavefunctions $\psi_1 $ and $\psi_2 $ according to
\begin{equation}  \label{ansatz1}
\begin{split}
\psi_1 \equiv  \psi_1 (t, r, \theta, \phi)   =  e^{i(\nu \phi - \omega t)}
\left(\begin{matrix}  \psi^{(1)}_1 ( r, \theta )  \\ \psi^{(2)}_1 ( r, \theta )  \\ \end{matrix}\right)
=  e^{i(\nu \phi - \omega t)}
\left(\begin{matrix}  -R_2 (r) S_1 (\theta )  \\  -R_1 (r) S_2 (\theta ) \\ \end{matrix}\right),  \\
\psi_2 \equiv  \psi_2 (t, r, \theta, \phi)   = e^{i(\nu \phi - \omega t)}
\left(\begin{matrix}  \psi^{(1)}_2 ( r, \theta )  \\ \psi^{(2)}_2 ( r, \theta )  \\ \end{matrix}\right)
=  e^{i(\nu \phi - \omega t)}
\left(\begin{matrix}  R_1 (r) S_1 (\theta )  \\  R_2 (r) S_2 (\theta ) \\ \end{matrix}\right),
\end{split}
\end{equation}
where $\omega$ and $\nu$ are  respectively  energy and  projection of the angular momentum of the spin $1/2$ particle. 
Note  that this factorization is not arbitrary, but is singled out by a demand of having separable equation of motion. Indeed, it gives rise to a straightforward separation of the equation of motion into radial and angular parts, as we show below.  

The first step in utilizing the factorization (\ref{ansatz1}), which includes  a separation of the azimuthal and time variables, gives rise to the set of two $2-$component equations
\begin{equation} \label{diracnovo6}
\begin{split}
\bigg[   \frac{i \omega r}{\sqrt{f}} \mathbb{1}   -  r\sqrt{f} \sigma_3 \partial_r  - \sigma_1 \partial_{\theta} 
- \frac{i \nu }{\sin \theta} \sigma_2   - \frac{1}{2} \cot \theta \sigma_1 
+  i \nu \frac{aqQ}{2r} \sqrt{f} \sigma_3  - \frac{iqQ}{ \sqrt{f}} \mathbb{1}  \bigg] \psi_2 (r, \theta) -  mr \mathbb{1}  \psi_1 (r, \theta) =0,\\
\bigg[ \frac{i \omega r}{\sqrt{f}} \mathbb{1}  +  r  \sqrt{f} \sigma_3 \partial_r  +  \sigma_1 \partial_{\theta}
+ \frac{i \nu }{\sin \theta} \sigma_2  + \frac{1}{2} \cot \theta \sigma_1 
-   i \nu   \frac{aqQ}{2r} \sqrt{f} \sigma_3   - \frac{iqQ}{ \sqrt{f}} \mathbb{1}  \bigg] \psi_1 (r, \theta )  - m r \mathbb{1}  \psi_2 (r, \theta ) =0.
\end{split}
\end{equation}
The second step, which involves a separation of the radial and polar angle variables, leads to the set of four coupled partial differential equations
\begin{equation} \label{diracnovo7}
\begin{split}
\frac{i \omega r}{\sqrt{f}} R_1 S_1   -  r\sqrt{f} ( \partial_r R_1 ) S_1 -  R_2 \partial_{\theta} S_2
- \frac{i \nu }{\sin \theta} (-i) R_2 S_2   - \frac{1}{2} \cot \theta R_2 S_2 &\\
+  i \nu \frac{aqQ}{2r} \sqrt{f}  R_1 S_1  +  mr R_2 S_1  - \frac{iqQ}{ \sqrt{f}} R_1 S_1    &=0, \\
\frac{i \omega r}{\sqrt{f}} R_2 S_2  +  r  \sqrt{f} ( \partial_r R_2 ) S_2  -R_1  \partial_{\theta} S_1
-\frac{i \nu }{\sin \theta} i R_1 S_1  - \frac{1}{2} \cot \theta R_1 S_1 &\\
-   i \nu   \frac{aqQ}{2r} \sqrt{f} R_2 S_2  + m r R_1 S_2  - \frac{iqQ}{ \sqrt{f}} R_2 S_2   &=0, \\
-  \frac{i \omega r}{\sqrt{f}} R_2 S_1   -  r\sqrt{f} ( \partial_r R_2 ) S_1  - R_1  \partial_{\theta} S_2
- \frac{i \nu }{\sin \theta} (-i) R_1 S_2   - \frac{1}{2} \cot \theta R_1 S_2 &\\
+  i \nu \frac{aqQ}{2r} \sqrt{f} R_2 S_1 -  mr R_1 S_1  + \frac{iqQ}{ \sqrt{f}} R_2 S_1    &=0, \\
- \frac{i \omega r}{\sqrt{f}} R_1 S_2  +  r  \sqrt{f}  (\partial_r  R_1) S_2 - R_2 \partial_{\theta} S_1
- \frac{i \nu }{\sin \theta} i R_2 S_1  - \frac{1}{2} \cot \theta R_2 S_1 &\\
-   i \nu   \frac{aqQ}{2r} \sqrt{f} R_1 S_2 - m r R_2 S_2  + \frac{iqQ}{ \sqrt{f}} R_1 S_2   &=0.
\end{split}
\end{equation}
After dividing above equations  respectively with $ R_2 S_1 $, $R_1 S_2$, $R_1 S_1$, $R_2 S_2$ one finds that this system of equations  is completely separable,
\begin{equation} \label{diracnovo8}
\begin{split}
\frac{i \omega r}{\sqrt{f}} \frac{R_1}{R_2}   -  r\sqrt{f} \frac{ \partial_r R_1 }{R_2} +  i \nu \frac{aqQ}{2r} \sqrt{f}  \frac{R_1}{R_2}  
- \frac{iqQ}{ \sqrt{f}} \frac{R_1}{R_2} +  mr =
\frac{\partial_{\theta} S_2}{S_1}  + \frac{ \nu }{\sin \theta} \frac{ S_2}{S_1}   + \frac{1}{2} \cot \theta \frac{ S_2}{S_1} \equiv \lambda,\\
\frac{i \omega r}{\sqrt{f}} \frac{R_2}{R_1}   +  r\sqrt{f} \frac{ \partial_r R_2 }{R_1} -  i \nu \frac{aqQ}{2r} \sqrt{f}  \frac{R_2}{R_1}  
- \frac{iqQ}{ \sqrt{f}} \frac{R_2}{R_1}  +  mr =
\frac{\partial_{\theta} S_1}{S_2}  - \frac{ \nu }{\sin \theta} \frac{ S_1}{S_2}   + \frac{1}{2} \cot \theta \frac{ S_1}{S_2} \equiv \lambda_1,
\\
- \frac{i \omega r}{\sqrt{f}} \frac{R_2}{R_1}   -  r\sqrt{f} \frac{ \partial_r R_2 }{R_1} +  i \nu \frac{aqQ}{2r} \sqrt{f}  \frac{R_2}{R_1}  
+ \frac{iqQ}{ \sqrt{f}} \frac{R_2}{R_1}  -  mr =
\frac{\partial_{\theta} S_2}{S_1}  + \frac{ \nu }{\sin \theta} \frac{ S_2}{S_1}   + \frac{1}{2} \cot \theta \frac{ S_2}{S_1} = \lambda,
\\
- \frac{i \omega r}{\sqrt{f}} \frac{R_1}{R_2}   +  r\sqrt{f} \frac{ \partial_r R_1 }{R_2} -  i \nu \frac{aqQ}{2r} \sqrt{f}  \frac{R_1}{R_2}  
+ \frac{iqQ}{ \sqrt{f}} \frac{R_1}{R_2}  -  mr =
\frac{\partial_{\theta} S_1}{S_2}  - \frac{ \nu }{\sin \theta} \frac{ S_1}{S_2}   + \frac{1}{2} \cot \theta \frac{ S_1}{S_2} = \lambda_1.
\end{split}
\end{equation}
Moreover,   it is easily seen that two separation constants $\lambda$ and $\lambda_1$, which have appeared in a process of separation are not mutually  independent, but subject to the requirement $\lambda  = -\lambda_1 $. In effect, the system of equations (\ref{diracnovo8}) gives rise to two angular equations
\begin{equation} \label{diracnovo-angular}
 \begin{split}  
&\partial_{\theta} S_2 + \frac{\nu}{\sin \theta} S_2 + \frac{1}{2} \cot \theta S_2 = \lambda S_1,  \\
& \partial_{\theta} S_1 - \frac{\nu}{\sin \theta} S_1 + \frac{1}{2} \cot \theta S_1 = -\lambda S_2,
\end{split}
\end{equation}
and two radial equations
\begin{equation} \label{diracnovo-radial}
\begin{split}  
\frac{i \omega r}{\sqrt{f}} R_1   - r\sqrt{f}  \partial_{r} R_1 +  i \nu \frac{aqQ}{2r} \sqrt{f} R_1
- \frac{iqQ}{ \sqrt{f}} R_1    & = \bigg(  \lambda - mr \bigg)  R_2,  \\
\frac{i \omega r}{\sqrt{f}} R_2   + r\sqrt{f}  \partial_{r} R_2 -  i \nu \frac{aqQ}{2r} \sqrt{f} R_2 
- \frac{iqQ}{ \sqrt{f}} R_2       &= - \bigg( \lambda + mr \bigg)  R_1,
\end{split}
\end{equation}
This system of radial equations can be used to study the behaviour of spinor quasinormal modes in the RN background.

This discussion we close with the analysis of the stability of chargeless, but massive fermionic modes.
For that purpose we recall that for the bosonic fields on  Kerr spacetime there exists a regime in which bosonic modes become unstable, due to superradiant growth \cite{chlee,gaina,detweiler_bossuperrad1, rosa_bossuperrad2, bossuperrad3, bossuperrad4, bossuperrad5}.
 Contrary to that,  in the same regime where the bosonic modes manifest instability, the fermionic fields on  Kerr spacetime under condition of extra slow rotation do not{\footnote{The absence of superradiance for Dirac field in a Kerr or Kerr-Newman background was proved for a first time in \cite{chlee}.}}, resulting in them
 being stable and  subject to a decay  only \cite{fermsuperrad1},\cite{fermsuperrad2},\cite{fermsuperrad3},\cite{fermsuperrad4},\cite{unruh}. 
In other words, unlike the equations of motion governing the bosonic fields, the single-particle Dirac equation is not subject to superradiance, and thus all modes decay
in that particular regime, which includes a setting where $mM \lesssim    ( l + \frac{1}{2}),$ as well as the limit of slow rotation, $\Omega / M << 1$\footnote{Note that $m$ is the mass of the perturbing field, $M$ is the mass of the black hole, $l$ is the orbital angular momentum number and $\Omega M$ is the angular momentum of a black hole.}.
Interestingly, fermionic fields on Schwarzschild or Reissner–Nordstr\" om spacetime  display  somewhat different characteristics, which makes them 
more susceptible of  exhibiting  the effect of superradiance and thus not  remaining stable.
These observations may be drawn by inspecting the bosonic and fermionic modes in question, either by inspecting the imaginary part of their bound state frequencies or by investigating the properties of the flux passing into the horizon  and the corresponding conservation law.  

Here we set to examine a possibility that a noncommutative deformation of RN spacetime, realized in a form of the effective metric 
(\ref{metric}), introduces certain changes to the above statements.
To start with, let us recall the form of  the wave function that solves the Dirac equation 
\begin{equation} \label{wec1}
\Psi = e^{i(\nu \phi - \omega t)} {\big( r^4 f \big) }^{-1/4} \left(\begin{matrix} \psi_1  \\ \psi_2  \\ \end{matrix}\right)  =
e^{i(\nu \phi - \omega t)} {\bigg( r^4  - 2Mr^3 + Q^2 r^2 \bigg) }^{-1/4} \left(\begin{matrix} -R_2(r) S_1(\theta)  \\ -R_1(r) S_2(\theta)  \\  R_1(r) S_1(\theta)  \\ R_2(r) S_2(\theta)  \\ \end{matrix}\right).
\end{equation}
Its  corresponding hermitian conjugate is defined as 
\begin{equation}
\begin{split}
\bar{\Psi} = -\Psi^{\dagger} \gamma^{0} = -i \Psi^{\dagger} 
\left( \begin{array}{ccccc}
  0  & \mathbb{1}  \\
   \mathbb{1}   & 0  \\ 
\end{array} \right),  
\end{split}
\end{equation}
and the covariant derivatives that include the spin connection part are given by
\begin{eqnarray}  \label{wec2}
\nabla_\mu \Psi  & = &  \partial_\mu  \Psi - \Gamma_\mu \Psi = \partial_\mu \Psi - \frac{1}{4}  \omega_{\mu}^{~~bc} \gamma_b \gamma_c \Psi,   \nonumber \\
\nabla_\mu  \bar{\Psi}  &  =  &  \partial_\mu \bar{\Psi} + \bar{\Psi} \Gamma_\mu =  \partial_\mu \bar{\Psi} +  \frac{1}{4}  \omega_{\mu}^{~~bc} \bar{\Psi} \gamma_b \gamma_c.    \nonumber
\end{eqnarray}
With these quantities at hand, one may define the stress-energy tensor as
\begin{equation} \label{wec3}
T_{\mu \nu} = \frac{i}{4} \Big[  \bar{\Psi} \gamma_{\mu} \nabla_{\nu} \Psi + \bar{\Psi} \gamma_{\nu} \nabla_{\mu} \Psi  - \big( \nabla_{\mu} \bar{\Psi} \big) \gamma_{\nu} \Psi - 
\big( \nabla_{\nu} \bar{\Psi} \big) \gamma_{\mu} \Psi  \Big].
\end{equation}

For the fermionic field on the Kerr spacetime the form of the solution for the wave function essentially (up to a different prefactor) has the same general form as (\ref{wec1}). The radial Dirac current and the corresponding conservation law in this case give rise  to the condition
\begin{equation}
\frac{dN}{dt} = {\bigg( {|R_1|}^2 -   {|R_2|}^2 \bigg)}_{r= r_h} \le 0,
\end{equation}
where $N$ is the number density and $r_h$ is the outer horizon radius, signalling the absence of superradiance \cite{Dolan:2015eua}.
For the case considered in this paper, i.e. deformed RN metric (\ref{metric}), the radial component of the Dirac current  $J^{\mu} = \bar{\Psi} \gamma^{\mu} \Psi = \bar{\Psi}  e_a^{~~\mu} \gamma^{a} \Psi  $ may be shown to have the form
\begin{equation}   \label{wec6}
\begin{split}
J^r &=  \bar{\Psi} \gamma^{r} \Psi = \bar{\Psi}  e_a^{~~r} \gamma^{a} \Psi  \\
&  = -i  {\bigg( r^4  - 2Mr^3 + Q^2 r^2 \bigg) }^{-1/2} \sqrt{f}  \left(\begin{matrix} R^{*}_1 S^{*}_1  & R^{*}_2 S^{*}_2  & - R^{*}_2 S^{*}_1  & - R^{*}_1 S^{*}_2  \\ \end{matrix}\right)
\left(\begin{matrix} 0 & 0 &  i &  0  \\ 0 & 0 & 0 & -i \\  -i & 0 & 0 & 0  \\ 0 & i & 0 & 0 \\ \end{matrix}\right)
\left(\begin{matrix} -R_2 S_1  \\ -R_1 S_2  \\  R_1 S_1  \\ R_2 S_2  \\ \end{matrix}\right)  \\
&  = \frac{1}{r^2}  \bigg( {|R_1|}^2 -   {|R_2|}^2 \bigg) \bigg( {|S_1|}^2 +   {|S_2|}^2 \bigg).
\end{split}
\end{equation}
At first glance, with the result  (\ref{wec6}) at hand, one might be led to think that the fermionic field on  deformed RN spacetime might also exhibit the absence of superradiance. This is due to the fact that the  integral of the radial current evaluated at the lower bound, i.e. at the outer horizon $r= r_h,$ has a natural interpretation as the flux passing into the horizon and the expression on the rhs of (\ref{wec6}) not being positive-definite.

However, a deeper inspection seems not to confirm this conclusion. In fact, a superradiance may be seen as a direct consequence of the second law of black hole thermodynamics. On the other hand, a crucial assumption that underlies the second law of black hole thermodynamics is that the condition  $T_{\mu \nu} t^{\mu} t^{\nu} \ge 0$ must hold for any time-like vector field, $t_{\mu} t^{\mu} > 0,$ with $T_{\mu \nu}$ being the stress-energy tensor. This condition is known as the weak energy condition.
Therefore, for the purpose of demonstrating   that a superradiance is absent in a given system, or at least that it doesn't show up for a particular choice of the system parameters, we would first have to make evident that there exists a sector in the parameter space  of that system where the weak energy condition is violated.
 In order to examine if the weak energy condition might possibly be violated within the context of fermionic field on  NC deformed RN background,
opening in this way a window for a possible violation of the second law of thermodynamics and a consequent loss of
the effect of superradiance within the same context, we take the time-like vector  $~ t^{\mu} \equiv e_0^{~~\mu} = ( e_0^{~~t}, e_0^{~~r}, e_0^{~~\theta}, e_0^{~~\phi}) = (\frac{1}{\sqrt{f}}, 0,0,0) ,~ t_{\mu} t^{\mu} > 0$, and first evaluate and then analyse  the bilinear form $T_{\mu \nu}  t^{\mu} t^{\nu} = T_{\mu \nu}   e_0^{~~\mu}  e_0^{~~\nu} 
 = T_{t t}   e_0^{~~t}  e_0^{~~t}.$  This gives
\begin{eqnarray}  \label{wec4}
T_{\mu \nu}  t^{\mu} t^{\nu}   & = &  \frac{i}{4} \Big[  \bar{\Psi} \gamma_{t} \nabla_{t} \Psi + \bar{\Psi} \gamma_{t} \nabla_{t} \Psi  - \big( \nabla_{t} \bar{\Psi} \big) \gamma_{t} \Psi - 
\big( \nabla_{t} \bar{\Psi} \big) \gamma_{t} \Psi  \Big]  e_0^{~~t}  e_0^{~~t}  \nonumber \\
&  =  & 2 \Bigg[ \frac{i}{4}   \bar{\Psi} \gamma_{t} \nabla_{t} \Psi  e_0^{~~t}  e_0^{~~t} -   \frac{i}{4} \big( \nabla_{t} \bar{\Psi} \big) \gamma_{t} \Psi  e_0^{~~t}  e_0^{~~t} \Bigg].
\end{eqnarray}
Inserting (\ref{metrictetrad}),(\ref{gammarep}) and (\ref{wec1}) into this equation leads to
\begin{eqnarray}  \label{wec5}
T_{\mu \nu}  t^{\mu} t^{\nu}   & = &  2\frac{{\bigg( r^4  - 2Mr^3 + Q^2 r^2 \bigg) }^{-1/2}}{4\sqrt{f}} \Bigg[ \omega \bigg( {|R_1|}^2 +   {|R_2|}^2 \bigg) \bigg( {|S_1|}^2 +   {|S_2|}^2 \bigg)  \nn\\
&& \hspace{4.6cm} + i \frac{Mr- Q^2}{2r^3} \bigg( {|R_1|}^2 -   {|R_2|}^2 \bigg) \bigg( {|S_1|}^2 +   {|S_2|}^2 \bigg) \Bigg] \nonumber \\
&&- 2\frac{{\bigg( r^4  - 2Mr^3 + Q^2 r^2 \bigg) }^{-1/2}}{4\sqrt{f}} \Bigg[- \omega \bigg( {|R_1|}^2 +   {|R_2|}^2 \bigg) \bigg( {|S_1|}^2 +   {|S_2|}^2 \bigg)  \nn\\
&& \hspace{4.8cm} + i \frac{Mr- Q^2}{2r^3} \bigg( {|R_1|}^2 -   {|R_2|}^2 \bigg) \bigg( {|S_1|}^2 +   {|S_2|}^2 \bigg)  \Bigg] \nonumber \\
&  = & \frac{{\bigg( r^4  - 2Mr^3 + Q^2 r^2 \bigg) }^{-1/2}}{2\sqrt{f}} 2 \omega \bigg( {|R_1|}^2 +   {|R_2|}^2 \bigg) \bigg( {|S_1|}^2 +   {|S_2|}^2 \bigg).
\end{eqnarray}
It is clear that  outside the outer horizon the expression under the square root is greater than zero. Moreover,  the expression (\ref{wec5}) as a whole  is strictly positive-definite, that is $T_{\mu \nu} t^{\mu} t^{\nu} \ge 0,$ implying that the noncommutative deformation of the Reissner–Nordstr\" om spacetime does not violate the weak energy condition for the fermionic field.
This in turn implies that the weak energy condition is not violated for the Dirac particle in a Reisner Nordstrom spacetime subject to a noncommutative deformation. Since the key assumption for the second law of black hole thermodynamics is not violated, the law continues to hold and the superradiance is expected to occur for the fermionic field in the spacetime described by the effective (deformed RN) metric,
contrary to a first naive impression obtained by considering the radial component of the Dirac current.
This result is different from the case of the fermionic field on Kerr spacetime  in the near horizon region $f(r) \rightarrow 0$ and in the superradiant\footnote{The term superradiant is here used because  in this regime bosonic fields rapidly grow in time, thus exhibiting a superradiance. Dirac fields though remain stable in this regime, as they are not superradiant there.} regime $\omega <  \nu \frac{\Omega}{2M r_h}$, where $\omega$ is the frequency of the mode and $\nu$ is its azimuthal number,  $\frac{\Omega}{2M r_h}$ is the angular frequency of the horizon  and $r_h$ is the horizon radius. In the latter case the weak-energy condition is violated for the Dirac field on Kerr spacetime and consequently the effect of superradiance is absent.
    In our future work we plan to use the results obtained in this paper in order to study  the  massless as well as the massive fermionic perturbations of RN black hole in the presence of spacetime noncommutativity.

\vspace{1cm}

\noindent{\bf Acknowledgment} We would like to thank Kumar Gupta, Tajron Juri\'{c} and
Ivica Smoli\'{c} for fruitful discussion and useful comments. This  research was supported by the Croatian Science
Foundation Project No. IP-2020-02-9614 {\it{Search for Quantum spacetime in Black Hole QNM spectrum and Gamma Ray Bursts.}} The work of M.D.C.
and N.K. is supported by project 451-03-9/2021-14/200162 of the Serbian Ministry of Education and
Science.  This work
is partially supported by ICTP-SEENET-MTP Project NT-03 ”Cosmology-Classical
and Quantum Challenges” in frame of the Southeastern European Network in Theoretical and Mathematical Physics and the COST action CA18108 {\it{Quantum gravity phenomenology in the multimessenger approach.}}

\appendix

\section{Effective metric from first order noncommutative duality-Calculation of the first order effective dual metric}

Here we show that the equation (\ref{extendedKG}) can be reversely engineered to yield 
the first order (in the deformation parameter $a$) 
 effective metric  (\ref{NCdsRN}). 

Equation (\ref{extendedKG}) can be symbolically written in terms of an extended Klein-Gordon operator, extended to include a coupling to a gauge field
\begin{equation} \label{KG1}
 \big( {\Box_{g'}} + {\mathcal{O}}(a) \big) \Phi \equiv \bigg( g'^{\mu \nu} \big( \nabla'_{\mu} - i A_{\mu}  \big )  \big( \nabla'_{\nu} - i A_{\nu}  \big) + {\mathcal{O}}(a) \bigg) \Phi= 0.
\end{equation}
Corrections  are included in ${\mathcal{O}}(a)$ that is a generic expression and it designates symbolically a whole set of correction  terms in the equation  (\ref{extendedKG}) that are induced by the noncommutativity and are therefore linear in NC parameter $a$. Likewise,  $\nabla'_{\mu}$ is a covariant derivative with respect to the metric $g'_{\mu \nu}$\footnote{$\nabla'_{\mu} A^{\nu} = \partial_{\mu} A^{\nu} + \Gamma'^{\nu}_{~\lambda \mu} A^{\lambda}$  and  $\nabla'_{\mu} A_{\nu} = \partial_{\mu} A_{\nu} - \Gamma'^{\lambda}_{~ \mu \nu} A_{\lambda}$.} (\ref{dsRN}) and  $\Box_{g'}$ is the Klein-Gordon operator for the metric $ g'_{\mu \nu}$.  Note that by switching off a noncommutativity by letting $a \longrightarrow 0,$ all corrections that scale with $a$ disappear, and the KG equation reduces
to
\begin{equation} \label{KG2}
{\Box_{g'}}  \Phi \equiv  g'^{\mu \nu} \big( \nabla'_{\mu} - i A_{\mu}  \big )  \big( \nabla'_{\nu} - i A_{\nu}  \big)   \Phi = \frac{1}{\sqrt{-g'}} (\partial_{\mu} -iA_{\mu}) \bigg( \sqrt{-g'} ~ g'^{\mu \nu} \big( \partial_{\nu} - i A_{\nu} ) \bigg)  \Phi = 0.
\end{equation}

At this stage one is  naturally led to ponder over a possiblity that the  terms in (\ref{extendedKG}) which scale linearly with the NC parameter $a$ can actually  be soaked up by the already present KG operator ${\Box_{g'}}$ to yield a KG operator ${\Box_{g}}$ with a redefined metric that has managed to absorb within itself  noncommutative features of the original problem. 
 More concisely, the question to be posed is if there exists a metric which is able to meet the requirement
\begin{equation} \label{KG3}
\begin{split}
 \big(  {\Box_{g'}} + {\mathcal{O}}(a) \big) \Phi & = \bigg( g'^{\mu \nu} \big( \nabla'_{\mu} - i A_{\mu}  \big )  \big( \nabla'_{\nu} - i A_{\nu}  \big) + {\mathcal{O}}(a) \bigg) \Phi
\\
&\equiv {\Box_{g}}  \Phi  =   g^{\mu \nu} \big( \nabla_{\mu} - i A_{\mu}  \big )  \big( \nabla_{\nu} - i A_{\nu}  \big)  \Phi
    = \frac{1}{\sqrt{-g}}
   (\partial_{\mu} -iA_{\mu}) \bigg( \sqrt{-g} ~ g^{\mu \nu} \big( \partial_{\nu} - i A_{\nu}  \big)  \bigg)  \Phi = 0.
\end{split}
\end{equation}
where $\nabla_{\mu}$ is  a covariant derivative with respect to the new, effective metric $g_{\mu \nu}$. We point out that the gauge potential did not change upon switching to a new setting  and rewriting dynamics of the system in terms of the effective metric.

In order to find the metric tensor which satisfies the requirement (\ref{KG3}), one may try  with the following ansatz
\begin{equation}\label{effectivemetric}
g_{\mu\nu}=\begin{pmatrix}
 f &0& 0 & 0\\
 0&-\frac{1}{f} &0 & g_{r \phi}\\
 0 &0& - r^2 & 0 \\
  0 & g_{r \phi}& 0 & - r^2 \sin \theta \\
\end{pmatrix},
\end{equation}
The novel nonvanishing entry  $g_{r \phi}$ is assumed to depend only on variables $r$ and $\theta,$ since we expect that $\partial_{t}$ and $\partial_{\phi} $ are Killing vectors for the  effective metric as well.  Moreover, it is assumed to be at least linear in NC parameter $a,$  $g_{r \phi} \sim {\mathcal{O}}(a)$ since the effective metric $g_{\mu \nu}$ has to reduce to the original RN metric  $g'_{\mu \nu}$ in the limiting case $a \longrightarrow 0$. The inverse of the metric tensor (\ref{effectivemetric}) has nonvanishing entries at the same places

\begin{equation}\label{inverseeffectivemetric}
g^{\mu\nu}=\begin{pmatrix}
 \frac{1}{f} &0& 0 & 0\\
 0&-f +\frac{f^2 g^2_{r \phi}}{f g^2_{r \phi} -  r^2 \sin^2 \theta} &0 & \frac{f g_{r \phi}}{f g^2_{r \phi} -  r^2 \sin^2 \theta}\\
 0 &0& - \frac{1}{ r^2 } & 0 \\
  0 &  \frac{f g_{r \phi}}{f g^2_{r \phi} -  r^2 \sin^2 \theta}  & 0 &  \frac{1}{f g^2_{r \phi} -  r^2 \sin^2 \theta} \\
\end{pmatrix},
\end{equation}
It can be seen that while off-diagonal elements have a leading correction term that is linear in $a,$ the diagonal elements $g^{rr}$ and $g^{\phi \phi}$ have a leading correction term that is quadratic in $a.$ 
Likewise, the determinant and the square-root of the determinant of the effective metric (\ref{effectivemetric}) have a leading correction term that is quadratic in the NC parameter $a,  ~~ \sqrt{-g} = r^2 \sin \theta + {\mathcal{O}}(a^2)$. These observations will have a crucial role in the subsequent analysis, whose aim is to deduce the metric $g_{\mu \nu},$ satisfying the requirement  (\ref{KG3}).

The form of the metric (\ref{effectivemetric}) dictates which terms are going to survive after the equation 
(\ref{KG3}) is written out explicitly

\begin{eqnarray}
{\Box_{g}}  \Phi  &=& \frac{1}{\sqrt{-g}}
(\partial_{\mu} -iA_{\mu}) \bigg( \sqrt{-g} ~ g^{\mu \nu} \big( \partial_{\nu} - i A_{\nu} \big) \bigg)  \Phi \nn\\
&=& \frac{1}{\sqrt{-g}} \Bigg[ (\partial_t - i A_t)\bigg( \sqrt{-g}g^{tt} \big( \partial_t - i A_t   \big)  \bigg) + 
   (\partial_r - i A_r)\bigg( \sqrt{-g}g^{rr} \big( \partial_r - i A_r   \big)  \bigg) \nn\\
&& +  (\partial_r - i A_r)\bigg( \sqrt{-g}g^{r \phi} \big( \partial_{\phi} - i A_{\phi}   \big)  \bigg)
+ (\partial_{\theta} - i A_{\theta})\bigg( \sqrt{-g}g^{\theta \theta} \big( \partial_{\theta} - i A_{\theta}   \big)  \bigg) \nn\\
&& +  (\partial_{\phi} - i A_{\phi})\bigg( \sqrt{-g}g^{\phi r} \big( \partial_r - i A_r  \big)  \bigg)
+  (\partial_{\phi} - i A_{\phi})\bigg( \sqrt{-g}g^{\phi \phi} \big( \partial_{\phi} - i A_{\phi}  \big)  \bigg)  \Bigg] \Phi .\nn
 \end{eqnarray}
Taking into account the fact that the gauge potential has only time component, one finds that the equation of motion (\ref{KG3}) further  boils down to
\begin{eqnarray}
 \frac{1}{f} \bigg[ \partial_t^2 \Phi -2i A_t \partial_t  \Phi  - A_t^2 \Phi \bigg] &+& \frac{1}{\sqrt{-g}} \bigg[ \partial_r \big( \sqrt{-g}g^{rr}   \big)  \bigg] \partial_r \Phi
+ g^{rr} \partial^2_r \Phi + \frac{1}{\sqrt{-g}} \bigg[ \partial_r \big( \sqrt{-g}g^{r \phi}   \big)  \bigg] \partial_{\phi} \Phi \nn\\
&+&  2g^{r \phi} \partial_r \partial_{\phi} \Phi + \frac{1}{\sqrt{-g}} \bigg[ \partial_{\theta} \big( \sqrt{-g}g^{\theta \theta}   \big)  \bigg] \partial_{\theta} \Phi + g^{\theta \theta} \partial^2_{\theta} \Phi  + g^{\phi \phi}  \partial^2_{\phi} \Phi = 0. \nn
\end{eqnarray}

Focusing only on terms in the above equation that are at most linear in $a,$ and stacking it up against the equation  (\ref{extendedKG}) leads to the following two relations:
\begin{equation} \label{KGconditions}
\begin{split}
  \frac{aqQ}{r^3} \bigg( \frac{MG}{r} -  \frac{G Q^2}{r^2}   \bigg) \partial_{\phi} \Phi  &=  \frac{1}{\sqrt{-g}} \bigg[ \partial_r \big( \sqrt{-g}g^{r \phi}   \big)  \bigg] \partial_{\phi} \Phi,
 \\
 \frac{aqQ}{r^2} f  \partial_r  \partial_{\phi} \Phi &=  2 g^{r \phi}   \partial_r \partial_{\phi} \Phi.
\end{split}
\end{equation}

The solution to this set of relations, which is consistent with the requirement $g^{r \phi} =\frac{f g_{r \phi}}{f g^2_{r \phi} -  r^2 \sin^2 \theta}, $ finally gives for the dual effective metric
\begin{equation} \label{metric}
  g_{\mu \nu} =
\left( \begin{array}{ccccc}
  f & 0  & 0  & 0 \\
   0   & -\frac{1}{f} & 0 & -\frac{aqQ}{2} \sin^2 \theta \\ 
   0  & 0 & -r^2  &  0 \\
   0  & -\frac{aqQ}{2} \sin^2 \theta & 0 & -r^2 \sin^2 \theta \\
\end{array} \right)  
\end{equation}

and for its inverse metric

\begin{equation}\label{InvMetric}
g^{\mu \nu} =
\left( \begin{array}{ccccc}
  \frac{1}{f} & 0  & 0 & 0  \\
   0   & -f  &  0  &  \frac{aqQ}{2 r^2} f    \\ 
   0  & 0 & -\frac{1}{r^2 } & 0   \\
   0   & \frac{aqQ}{2 r^2} f  &  0 &  -\frac{1}{r^2  \sin^2 \theta } \\
\end{array} \right).
\end{equation}
Note that we demand $g_{\mu \nu} g^{\nu \rho} =  \delta_{\mu}^{~~\rho} + {\mathcal{O}}(a^2).$

We have thus shown that the equation of motion for a charged NC scalar field in a classical RN background, coupled  to  NC $U(1)$ gauge field
may be rewritten in terms of the equation of motion governing behaviour of a charged commutative scalar field (having the same charge $q$ as its NC counterpart), propagating in a modified RN geometry
\begin{equation} 
  {\rm d}s^2 = \Big(1-\frac{2MG}{r}+\frac{Q^2G}{r^2} \Big) {\rm d}t^2 - \frac{{\rm
d}r^2}{1-\frac{2MG}{r}+\frac{Q^2G}{r^2}} - aqQ \sin^2 \theta {\rm d} r {\rm d} \phi - r^2({\rm d}\theta^2 + \sin^2\theta{\rm d}\phi^2 ).
\end{equation}

\section{More general choices of the twist}

In this Appendix we briefly discuss more general forms of the twist operator, leading to more general NC deformations. This discussion if far from being complete, the detailed analysis we postpone for our future research.

\begin{enumerate}

\item 
For a deformation with a Killing twist operator, an arbitrary static, spherically symmetric metric 
\begin{equation}
{\rm d}s^2 = f(r) {\rm d}t^2 - \frac{{\rm
d}r^2}{f(r)} - r^2({\rm d}\theta^2 + \sin^2\theta{\rm d}\phi^2)
. \label{dsf}
\end{equation}
and an arbitrary static, spherically symmetric electromagnetic potential $A_\mu {\rm d}x^\mu = A(r){\rm d}t$, with arbitrary functions $f(r)$ and $A(r)$, our results continue to be valid. More precisely, from  equations (\ref{KGconditions}) it follows that the component $g^{r\varphi}$ is related to the function $f(r)$, while the radial derivative of $g^{r\varphi}$ is related to the radial derivative of $f(r)$. Note that in the RN case, the term $\frac{MG}{r} -  \frac{G Q^2}{r^2}$ is nothing else but $\frac{r}{2} \partial_r f$.  Therefore, the component $g^{r\varphi}$ will have the same form as in (\ref{InvMetric}) with an arbitrary function $f(r)$ from (\ref{dsf}). On the other hand, in this more general case the only nonvanishing component of the field strength tensor is $F_{tr} = -\partial_r A(r)$. We conclude that the effective metric will retain the same general form (\ref{metric}) with
$~g_{r \phi} = a\frac{r^2}{2} F_{tr} \sin^2 \theta, ~$ and its inverse metric will have a different $g^{r\varphi}$ component given by $g^{r\varphi} = -\frac{a}{2} F_{tr} f(r)= \frac{a}{2}f(r)\partial_r A(r)$.

\item
For a definition with a semi-Killing twist operator, such as 
\begin{equation}
\mathcal{F} =  e^{-\frac{ia}{2} (\partial_t\otimes\partial_r - \partial_r\otimes\partial_t)} \nn
\end{equation}
for the RN metric, the Seiberg-Witten expanded (up to first order) action for the NC scalar field is given by
\begin{eqnarray}
S &=& \int
\d^4x \Big( \sqrt{-g}\, \big( g^{\mu\nu}D_\mu\phi^+D_\nu\phi -\mu^2\phi^+\phi +\frac{\mu^2}{2}\theta^{\alpha\beta}F_{\alpha\beta}\phi^+\phi \nonumber\\
&& + \frac{1}{2}\theta^{\alpha\beta}g^{\mu\nu}\big( -\frac{1}{2}D_\mu\phi^+F_{\alpha\beta}
D_\nu\phi +(D_\mu\phi^+)F_{\alpha\nu}D_\beta\phi + (D_\beta\phi^+)F_{\alpha\mu}D_\nu\phi\big) \big)\nn\\
&&-\frac{i}{2}\theta^{\alpha\beta}\partial_\alpha \big(\sqrt{-g}g^{\mu\nu}\big) D_\mu\phi^+D_\beta D_\nu\phi \big) \Big)
,\nn 
\end{eqnarray} 
where $ D_{\nu} \phi = (\partial_{\nu} - i A_{\nu}) \phi. $
We notice that an additional (compared to (3.34) in \cite{Ciric:2017rnf}) term $-\frac{i}{2}\theta^{\alpha\beta}\partial_\alpha \big(\sqrt{-g}g^{\mu\nu}) D_\mu\phi^+ D_\beta D_\nu\phi$ will lead to an equation of motion for the field $\phi$ that is third order in derivatives. This immediately signals that equation cannot be reduced to an equation of motion for a commutative scalar filed in an effective metric, since that equation is necessarily a 2nd order differential equation.

\item
For a completely arbitrary (within a physical reason) well defined twist there is no guaranty that the first order duality will hold. Moreover, based on the results for the semi-Killing twist deformation it is very likely that there will be no duality between the propagation of the NC scalar field on a fixed background and the propagation of commutative scalar field in an effective background.

\end{enumerate}

From this brief analysis we can conclude that the deformation by a Killing twist is a special one and it is the only one that corresponds to the semi-classical approximation we use in our work.

\end{document}